\documentclass[10pt,twocolumn,conference]{IEEEtran}

\usepackage{graphicx,dsfont,amssymb,mathrsfs,amsmath,enumerate,amsfonts,hyperref}

\newcommand{\bbE}{\mathbb E}
\newcommand{\bbR}{{\mathbb R}}

\newcommand{\calK}{{\mathcal K}}
\newcommand{\calM}{{\mathcal M}}
\newcommand{\calQ}{{\mathcal Q}}

\newcommand{\calX}{{\mathcal X}}
\newcommand{\calW}{{\mathcal W}}

\newcommand{\bfX}{{\mathbf X}}
\newcommand{\bfY}{{\mathbf Y}}
\newcommand{\bfW}{{\mathbf W}}\newcommand{\bfw}{{\mathbf w}}
\newcommand{\bfp}{{\mathbf p}}
\newcommand{\bfal}{\mbox{\boldmath $\alpha$}}

\newcommand{\scrP}{\mathscr P}
\newcommand{\scrW}{\mathscr W}

\newcommand{\Bernoulli}{\mathrm{Bernoulli}}

\newtheorem{theorem}{Theorem}[section]

\newtheorem{lemma}[theorem]{Lemma}

\newtheorem{definition}[theorem]{Definition}
\newtheorem{conjecture}[theorem]{Conjecture}

\begin{document}

% paper title
\title{Saddle-point Solution of the Fingerprinting Capacity Game Under the Marking Assumption}

\author{
\authorblockN{Yen-Wei Huang}
\authorblockA{Beckman Inst., Coord. Sci. Lab and ECE Department\\
University of Illinois at Urbana-Champaign, USA\\
Email: huang37@illinois.edu}
\and
\authorblockN{Pierre Moulin}
\authorblockA{Beckman Inst., Coord. Sci. Lab and ECE Department\\
University of Illinois at Urbana-Champaign, USA\\
Email: moulin@ifp.uiuc.edu}
}

% make the title area
\maketitle

\begin{abstract}
We study a fingerprinting game in which the collusion channel is unknown. The encoder embeds fingerprints into a host sequence and provides the decoder with the capability to trace back pirated copies to the colluders.

Fingerprinting capacity has recently been derived as the limit value of a sequence of maxmin games with mutual information as the payoff function. However, these games generally do not admit saddle-point solutions and are very hard to solve numerically. Here under the so-called Boneh-Shaw marking assumption, we reformulate the
capacity as the value of a single two-person zero-sum game, and show that it is achieved by a saddle-point solution.

If the maximal coalition size is $k$ and the fingerprint alphabet is binary, we derive equations that can numerically solve the capacity game for arbitrary $k$. We also provide tight upper and lower bounds on the capacity. Finally, we discuss the asymptotic behavior of the fingerprinting game for large $k$ and practical implementation issues.
\end{abstract}

\section{Introduction}

Fingerprinting is a technique for copyright protection. It was first proposed by Wagner in 1983 \cite{Wagner1983} and has drawn a lot of attention in recent years. The content distributor embeds a unique mark, or \emph{fingerprint}, within each licensed copy. By forming a group of users (\emph{pirates}), the \emph{coalition} can detect the fingerprints by inspecting the marks in each copy, and create a \emph{forgery} that has only weak traces of their copies. A collusion-resistant fingerprinting system is designed to combat the collusive attacks.

Boneh and Shaw in \cite{Boneh1998} proposed the \emph{marking assumption} for the fingerprinting problem. In this setup, fingerprints are a string of marks allocated throughout the host content. The locations of the marks are assumed unknown to the pirates. By comparing their available copies, the coalition can remove or replace the detected marks, but cannot modify those marks at which their copies agree. As a result, we can ignore the host sequence and consider only the fingerprints in our analysis.

Tardos in 2003 \cite{Tardos2003} invented a simple but efficient randomized fingerprinting code that invites many subsequent works, such as \cite{vSkori'c2008, Furon2008}. Amiri and Tardos recently \cite{Amiri2009} (and independently of our work) further improved the rate by constructing a code based on a two-person zero-sum game. Although the code is far more efficient than the previous scheme, the intense computational complexity makes it less appealing for practical use.

A few researchers have also studied the problem from the information-theoretic point of view \cite{Amiri2009, Moulin2003, Moulin2008, Moulin2008a, Anthapadmanabhan2008}. Here we focus on finding the maximum achievable rate, or \emph{capacity}, of the fingerprinting system. Recently, Moulin in \cite{Moulin2008a} provided the capacity formula in a general setup. We study specifically the marking assumption in this paper and show that the capacity is indeed the rate achieved in \cite{Amiri2009}.

One concern is that neither the encoder nor the decoder knows the actual coalition size in real applications \cite{Furon2009}. We show that this is actually not a big issue. The saddle-point property states that for a fingerprinting code designed for a maximal coalition size $k$, there exists a unique saddle-point solution that achieves the capacity. That is, neither the content distributor nor the coalition can gain by deviating from its optimal strategy. As a result, the system is secure for any collusive attack of size no more than $k$. Furthermore, even if the size-$k$ anticipation is violated, no innocent user is accused \cite{Moulin2008a}. The pirates are simply too powerful and we have not enough evidence to accuse them. Instead, the decoder gives us the more probable suspects which may allow the legal authority to do further investigation.

In this paper, we reformulate the capacity formula in \cite{Moulin2008a} as the value of a single two-person zero-sum game and show that it admits a saddle-point solution. In the binary alphabet case, new capacity bounds are provided. The proofs not only show that the binary fingerprinting capacity is in $\Theta(1/{k^2})$, but they also provide secure strategies for both players of the game. Along with the numerical saddle-point solutions for small $k$, we study the asymptotic behavior of the game for large $k$.

The outline of the paper is as follows: In Section \ref{sec:def}, we formally define fingerprinting capacity and review the capacity formula derived in \cite{Moulin2008a}. The derivation of the single fingerprinting capacity game is shown in Section \ref{sec:game}, and Section \ref{sec:bin} is devoted to the binary alphabet case.

\section{Problem Statement} \label{sec:def}

\subsection{Notation}

We use capital letters to represent random variables, and lowercase letters to their realizations. Boldfaces denote vectors, and calligraphic letters denote sets. For example, $\bfX \in \calX^n$ denotes a random vector $(X_1, \ldots, X_n)$, with each $X_i$ taking values in $\calX$. The probability distribution of $\bfX$ is denoted by $p_\bfX$. The entropy of a random variable $X$ is denoted by $H(X)$. The mutual information of $X$ and $Y$, with joint pmf $p$ is denoted by $I_p(X;Y) = H(X) - H(X|Y)$. We also denote the binary entropy function by $h_2(p) \triangleq -p\log p - (1-p)\log (1-p)$ and $h_2(\bfp) = \left(h_2(p_1), \ldots, h_2(p_n)\right)'$. The KL divergence between two Bernoulli random variables with expectations $p$ and $q$ is denoted by $d_2(p\|q) \triangleq p\log\frac{p}{q}+(1-p)\log\frac{1-p}{1-q}$. $\log$ denotes base 2 logarithm and $\ln$ denotes natural logarithm. Mathematical expectation is denoted by the symbol $\bbE$. The shorthands $f \sim g$ and $f \gtrsim g$ denote asymptotic relations $\lim_{k \rightarrow \infty} \frac{f(k)}{g(k)} = 1$ and $\liminf_{k \rightarrow \infty} \frac{f(k)}{g(k)} \geq 1$ respectively.

\subsection{Overview}

Let $\calQ = \{0,1,\ldots,q-1\}$ denote a size-$q$ fingerprint alphabet, and $\calM = \{1,\ldots,m\}$ denote the set of user indices. The fingerprint encoder assigns each user a length-$n$ fingerprint, using an encoding function
\begin{equation}
f_n: \calM \times \calW_n \rightarrow \calQ^n,
\end{equation}
where the secret key $W_n \in \calW_n$ is a random variable whose realization is known to the encoder and the decoder, but unknown to the pirates.

A coalition $\calK$ is any size-$k$ subset of $\calM$, and $\bfX_\calK = \{\bfX_1, \ldots, \bfX_k\}$ are the fingerprints available to the coalition. The \emph{collusion channel} produces the forgery $\bfY \in \calQ^n$ according to distribution $p_{\bfY|\bfX_\calK}$. The marking assumption states that if for some $j \in \{1, \ldots, n\}$, $x_{1,j} = \cdots = x_{k,j}$, then $y_j = x_{1,j}$.

Not knowing the actual collusion channel $p_{\bfY|\bfX_\calK}$, the single-output decoder
\begin{equation}
g_n: \calQ^n \times \calW_n \rightarrow \calM
\end{equation}
accuses exactly one user based on the forgery $\bfY$ and the secret key $W_n$. The encoding and decoding functions $f_n$ and $g_n$ are deterministic, but a fingerprinting code is a random variable $(F_n, G_n)$ whose distribution is characterized by that of $W_n$. Under fingerprinting code $(F_n, G_n)$, the worst-case error probability is defined as
\begin{equation}
P^*_e(F_n, G_n,k) = \max_{\substack{\calK \subseteq \calM\\ |\calK| \leq k}
}~\max_{p_{\bfY|\bfX_\calK}} Pr\left(G_n(\bfY,W_n)
\notin \calK\right),
\end{equation}
where the second maximization is over all $p_{\bfY|\bfX_\calK}$ satisfying the marking assumption.

\subsection{Fingerprinting Capacity}

We now define fingerprinting capacity and review the capacity formula \cite{Moulin2008a} under the marking assumption. Capacity is achieved using a random coding scheme.

\begin{definition}
A rate $R$ is achievable for the $q$-ary alphabet and size-$k$ coalitions if there exists a sequence of fingerprinting codes $(F_n, G_n)$ for $m = \lceil 2^{nR} \rceil$ users such that
\begin{equation}
\lim_{n \rightarrow \infty} P^*_e(F_n, G_n,k) = 0.
\end{equation}
\end{definition}

\medskip

\begin{definition}
Fingerprinting capacity $C_{k,q}$ is the supremum of all achievable rates for the $q$-ary alphabet and size-$k$ coalitions.
\end{definition}

\medskip

Now for a random variable $W$ defined over an alphabet $\calW = \left\{1, 2, \ldots, l\right\}$, we define the embedding class
\begin{equation}
\scrP^l_{X_{\calK} W} = \left\{p_{X_{\calK}W} (x_\calK, w) = p_W(w) \prod_{i=1}^k p_{X|W}(x_i|w) \right\},
\end{equation}
the collusion class
\begin{eqnarray}\label{eqn:collusion}
\lefteqn{\scrP_{Y|X_\calK} = \{ p_{Y|X_\calK}: p_{Y|X_{\pi(\calK)}} = p_{Y|X_\calK}, \forall \pi;} \nonumber\\
&&p_{Y|X_\calK}(y|x_\calK) = 1 \textrm{ if } y = x_1 = \cdots = x_k \},
\end{eqnarray}
where $\pi: \calK \rightarrow \calK$ is a permutation of the coalition $\calK$, and the function
\begin{equation}
C^l_{k,q} = \frac{1}{k} \max_{p_{X_\calK W} \in
\scrP^l_{X_\calK W}} ~ \min_{p_{Y|X_\calK} \in
\scrP_{Y|X_\calK}} ~
I(X_\calK;Y|W).\label{eqn:Conel}
\end{equation}

\begin{theorem} \cite[Theorem 3.4]{Moulin2008a}
The fingerprinting capacity $C_{k,q}$ for the $q$-ary alphabet and size-$k$ coalitions is
\begin{equation}
C_{k,q} = \lim_{l \rightarrow \infty}
C^l_{k,q}.
\end{equation}
\end{theorem}

\medskip

Fingerprinting capacity is the limit value of a sequence of maxmin games. For any fixed $l$, $C^l_{k,q}$ is the maxmin value of a two-person zero-sum game with the content distributor as the maximizer and the coalition as the minimizer. In the achievability proof, $W$ is a time-sharing random variable. As $l$ increases, it
gives the content distributor more flexibility in choosing the codes. Hence the sequence $C^l_{k,q}, 1 \leq l \leq \infty$, is nondecreasing and admits a finite limit.

However, it is not an easy task to evaluate $C^l_{k,q}$ as well as the capacity-achieving probability distributions, even for small values of $l$. The reason is that a saddle-point solution is generally not guaranteed. For the binary alphabet ($q = 2$) and $l = 1$, we can derive that
$$C^1_{k,2} = \frac{1}{k}2^{-(k-1)},$$
which is not achieved by a saddle-point solution when $k>2$. Also, this is very loose lower bound for $C_{k,2}$ comparing to the $\Theta(k^{-2})$ bound we will show in Sec. \ref{subsec:bounds}.

\section{The Two-person Zero-sum Game of Fingerprinting Capacity} \label{sec:game}

To establish the desired saddle-point property, we first reformulate the fingerprinting capacity as the value of a single maxmin game. Consider an auxiliary random vector $\bfW$ drawn from the $q$-dimensional probability simplex
\begin{equation}
\scrW^q \triangleq \left\{ \bfw \in \bbR^q: \sum_{x=0}^{q-1} w_x = 1 \textrm{ and } 0 \leq w_x \leq 1, x \in \calQ \right\}
\end{equation}
and the class of joint distributions
\begin{eqnarray}\label{eqn:jointdist}
\scrP_{X_{\calK} \bfW} = \{ p_{X_{\calK}\bfW} (x_\calK, \bfw) = p_\bfW(\bfw) \prod_{i = 1}^k p_{X|\bfW}(x_i|\bfw), \nonumber\\
\textrm{where } p_{X|\bfW}(x|\bfw) = w_x, x \in \calQ \}.
\end{eqnarray}
Then we can express $C_{k,q}$ as in the following theorem.

\medskip

\begin{theorem} \label{thm:maxmin}
\begin{equation}\label{eqn:maxmin}
C_{k,q} = \frac{1}{k} \max_{p_{X_\calK \bfW} \in
\scrP_{X_\calK \bfW}} ~ \min_{p_{Y|X_\calK} \in
\scrP_{Y|X_\calK}} ~ I(X_\calK;Y|\bfW).
\end{equation}
\end{theorem}

\medskip

\begin{proof}
Note that the class $\scrP_{X_{\calK} \bfW}$ is compact and the payoff function is bounded, hence the maximizer exists. Denote the right-hand side of \eqref{eqn:maxmin} by $C'_{k,q}$. We can show that $C'_{k,q} \geq C_{k,q}$ and $C'_{k,q} \leq C_{k,q}$ respectively. For lack of space we skip the complete proof but give only the outline. For any finite $l$, let
\begin{equation} \label{eqn:prob_l}
p^l_{X_\calK W} (x_\calK, w) = p^l_W (w) \prod_{i \in \calK} p^l_{X|W} (x_i|w) \in \scrP^l_{X_\calK W}
\end{equation}
and $p^l_{Y|X_\calK} \in \scrP_{Y|X_\calK}$ be the probability distributions that achieve \eqref{eqn:Conel}. Let
\begin{equation}
p_\bfW(\bfw) = \sum_{w \in S_\bfw} p^l_W(w),
\end{equation}
where
\begin{equation}
S_\bfw = \left\{ w \in \calW: p^l_{X|W}(x|w) = w_x, x \in \calQ \right\},\quad \bfw \in \scrW^q,
\end{equation}
then we can verify that the resulting $p_{X_\calK \bfW}$ satisfies
\begin{equation} \label{eqn:eqiv}
I_{p_{X_\calK \bfW},p^l_{Y|X_\calK}}(X_\calK; Y|\bfW) = I_{p^l_{X_\calK W},p^l_{Y|X_\calK}}(X_\calK; Y|W).
\end{equation}

\eqref{eqn:eqiv} shows that for any $p^l_{X_\calK W}$ defined in \eqref{eqn:prob_l}, we can find a probability distribution in $\scrP_{X_{\calK} \bfW}$ that achieves $C_{k,q}^l$. Thus $C'_{k,q} \geq C_{k,q}$.

The proof of $C'_{k,q} \leq C_{k,q}$ utilizes the continuity property of mutual information, by which we can show that the sequence $\langle C^l_{k,q} \rangle^\infty_{l=1}$ is lower bounded by a sequence converging to $C'_{k,q}$. Hence $C'_{k,q} = C_{k,q}$.
\end{proof}

\medskip

Theorem \ref{thm:maxmin} states the fingerprinting capacity as the maxmin value of a two-person zero-sum game. Note that since $p_{X|\bfW}$ is actually fixed in the class of joint distributions defined in \eqref{eqn:jointdist}, the maximizer only has control over $p_\bfW$, which lies within the class of probability distributions over $\scrW^q$, denoted by $\scrP_\bfW$. Also, the payoff function $I(X_\calK;Y|\bfW)$ is a linear function of $p_\bfW$ for fixed $p_{Y|X_\calK}$ and a convex function of $p_{Y|X_\calK}$ for fixed $p_\bfW$. By the minimax theorem \cite{Sion1958}, the game admits a saddle-point solution. In the game-theoretic point of view, this is a so-called convex game \cite[$\S2.5$]{Petrosjan1996}. The maximizer has an optimal mixed-strategy with a finite support and the minimizer has an optimal unique pure-strategy. Furthermore, the maxmin value equals the minmax value of the same game restricting both players with pure strategies. The following theorem states these properties.

\begin{theorem}\label{thm:saddle_point}
\begin{eqnarray}
C_{k,q} &=& \frac{1}{k} \min_{p_{Y|X_\calK} \in \scrP_{Y|X_\calK}} ~
\max_{\bfw \in \scrW^q} ~ I(X_\calK;Y|\bfW = \bfw)\nonumber\\
&=& \frac{1}{k} \max_{p_\bfW \in \scrP_\bfW} ~ \min_{p_{Y|X_\calK} \in \scrP_{Y|X_\calK}} ~ I(X_\calK;Y|\bfW).
\label{eqn:minmax}
\end{eqnarray}
\end{theorem}

\medskip

\section{Capacity for The Binary Alphabet} \label{sec:bin}

We have established the existence of a saddle-point solution for the capacity game. For the rest of the paper, we focus on the binary alphabet case and see how the game can be solved. 

\subsection{Game Definition}

We can simplify the game as follows:
\begin{enumerate}
\item {\bf Fingerprinting Code.} $\calQ = \{0,1\}$. The auxiliary random vector $\bfW$ now has only one degree of freedom, and we redefine it as $W \in [0,1]$. $p_W$ denotes its distribution, and $\scrW_S$ the support of $p_W$. $p_{X |W} \sim \Bernoulli(W)$ is fixed.
\item {\bf Collusion Channel.} Since $p_{Y|X_\calK} \in \scrP_{Y|X_\calK}$ defined in \eqref{eqn:collusion} is invariant to permutations of $\calK$, it takes the form $p_{Y|Z}$, where $Z \triangleq \sum_{i = 1}^k X_i \in \left\{0, 1, \ldots, k\right\}$ is the number of 1's in $X_\calK$. Let $\bfp = (p_0, \ldots, p_k)'$, where $p_z \triangleq p_{Y|Z}(1|z), z = 0, \ldots, k$. The marking assumption enforces that $p_0 = 0$ and $p_k = 1$, and the collusion channel is then completely characterized by $\bfp$.
\item {\bf Capacity.} If we let $\bfal(w) = \left(\alpha_0(w), \ldots, \alpha_k(w)\right)'$, where
\begin{equation}\label{eqn:alpha}
\alpha_z(w) \triangleq p_{Z|W}(z|w) = \binom{k}{z} w^z (1-w)^{k-z}
\end{equation}
is the binomial distribution with parameter $w \in [0,1]$, then we have
\begin{eqnarray}
\hat{C}(w,\bfp) &\triangleq& I(X_\calK; Y|W = w) \nonumber \\
&=& H(Y|W = w) - H(Y|X_\calK, W = w)\nonumber\\
&=& h_2\left(\sum_{z=0}^{k} p_z \alpha_z(w)\right) - \sum_{z=0}^{k} h_2(p_z) \alpha_z(w)\nonumber\\
&=& h_2(\bfal'\bfp) - \bfal' h_2(\bfp).\label{eqn:C_hat}
\end{eqnarray}
Another representation of $\hat{C}(w,\bfp)$ is
\begin{eqnarray}
\hat{C}(w,\bfp) &=& D(p_{YZ|W} \| p_{Y|W}p_{Z|W} | W = w) \nonumber\\
&=& \sum_{z=0}^{k} \alpha_z(w)~d_2(p_z \| \bfal'\bfp)
\end{eqnarray}

The fingerprinting capacity game for the binary alphabet under the marking assumption can then be written as
\begin{eqnarray}
C_{k,2} &=& \frac{1}{k} \max_{p_W} ~ \min_{\bfp} ~ \bbE_{p_W}\left[\hat{C}(W,\bfp)\right]\label{eqn:C_k2_maxmin}\\
&=& \frac{1}{k} \min_{\bfp} ~ \max_{w} ~ \hat{C}(w,\bfp)\label{eqn:C_k2_minmax}.
\end{eqnarray}
\end{enumerate}

\subsection{Analysis of the Convex Game} \label{subsec:analysis}

\begin{lemma}\label{thm:symmetry}
If $\bfp^*$ is the minimizer in \eqref{eqn:C_k2_maxmin} and \eqref{eqn:C_k2_minmax}, then
\begin{equation}\label{eqn:psym}
p^*_z = 1-p^*_{k-z}, z=0, \ldots, k.
\end{equation}
Also, if $p^*_W$ is the maximizer of \eqref{eqn:C_k2_maxmin}, then
\begin{equation}
p^*_W(w) = p^*_W(1-w), \forall w \in [0,1].
\end{equation}
\end{lemma}

\medskip

We skip the complete proof of Lemma \ref{thm:symmetry} here but only explain its idea. Note that $p^*_z$ represents the probability of assigning $Y$ as $1$ when $X_\calK$ has $z$ 1's and $(k-z)$ 0's. By symmetry we should expect in colluders' capacity-achieving strategy, the probability of assigning $Y$ as $0$ when $X_\calK$
has $(k-z)$ 1's and $z$ 0's to also be $p^*_z$, i.e., $p^*_{k-z} = 1-p^*_z, z = 0,\ldots,k$. Similarly, the capacity-achieving fingerprinting codes should have the same distribution for 0 and 1, hence $p^*_W$ should be symmetric as stated.

Owing to the existence of the saddle-point solution, $\bfp^*$ and $p^*_W$ must satisfy the following:
\begin{enumerate}
\item When $\bfp = \bfp^*$ is fixed, $\hat{C}(w,\bfp^*)$ is a differentiable function over the unit interval. The support of $p^*_W$, $\scrW^*_S$, can only take values at the maximizers of $\hat{C}(w,\bfp^*)$. Hence we have
\begin{equation} \label{eqn:dcdw}
\left\{\begin{array}{l}
\hat{C}(w,\bfp^*) = kC_{k,2}\\
\frac{\partial}{\partial w} \hat{C}(w,\bfp^*) = 0
\end{array}\right., \quad \forall w \in \scrW_S^*.
\end{equation}

\item When $p_W = p^*_W$ is fixed, and if we only consider $\bfp$ that satisfies \eqref{eqn:psym}, then we have
\begin{equation} \label{eqn:dcdp}
\bbE_{p^*_W}\left[\frac{\partial}{\partial p_z}\hat{C}(W,\bfp^*)\right] = 0, \quad z = 1,\ldots, \left\lfloor \frac{k-1}{2} \right\rfloor.
\end{equation}
\end{enumerate}

By the convexity in $\bfp$ of the payoff function, we know that $|\scrW^*_S| \leq \left\lfloor\frac{k+1}{2}\right\rfloor$ (see \cite[$\S2.5$]{Petrosjan1996}). With a fixed support cardinality, we can obtain candidate capacity-achieving distributions $\bfp^*$ and $p^*_W$ by solving \eqref{eqn:dcdw} and \eqref{eqn:dcdp}, and then verify those candidate distributions are optimal by examining the second partial derivatives. Once $\bfp^*$ and $p^*_W$ are found, we can get $C_{k,2}$ by substituting them into \eqref{eqn:C_k2_maxmin}.

\subsection{Bounds on Capacity} \label{subsec:bounds}

For general $k$, the following two theorems gives us $C_{k,2} =
\Theta(1/k^2)$.

\begin{theorem}\label{thm:ub}
\begin{equation}
C_{k,2} \leq \frac{1}{k^2 \ln 2} = \frac{1.443\ldots}{k^2}.
\end{equation}
\end{theorem}

\medskip

\begin{proof}

Consider the so-called ``interleaving
attack'' $\bfp^\infty$ defined by
$$p^\infty_z = \frac{z}{k}, \quad z = 0, \ldots, k,$$
then we have
\begin{eqnarray}
C_{k,2} &=& \frac{1}{k} \min_{\bfp} ~ \max_{w} ~
\hat{C}(w,\bfp)\nonumber\\
&\leq& \frac{1}{k} \max_{w}~\hat{C}(w,\bfp^\infty)\nonumber\\
&=& \frac{1}{k} \max_{w}~ \left\{h_2(w)-\sum_{z=0}^{k}\alpha_z(w)h_2\left(\frac{z}{k}\right)\right\}\nonumber\\
&\leq& \frac{1}{k^2 \ln 2},
\end{eqnarray}
where the last inequality results from \cite[Theorem
4.3]{Anthapadmanabhan2008}.
\end{proof}

\begin{theorem}\label{thm:lb}
\begin{equation}
C_{k,2} \geq \frac{2}{k^2 \pi^2\ln 2} = \frac{0.292\ldots}{k^2}.
\end{equation}
\end{theorem}

\medskip

\begin{proof}
Consider the continuous distribution
\begin{equation}\label{eqn:pwinfty}
p^\infty_W(w) = \frac{1}{\pi \sqrt{w(1-w)}}, w \in (0,1),
\end{equation}
then we have
\begin{eqnarray*}
C_{k,2} &=& \frac{1}{k} \max_{p_W} ~ \min_{\bfp} ~ \bbE_{p_W}\left[\hat{C}(W,\bfp)\right] \\
&\geq& \frac{1}{k} \min_{\bfp} ~ \bbE_{p^\infty_W}\left[\hat{C}(W,\bfp)\right]\\
&=& \frac{1}{k} \int_0^1 \sum_{z=0}^{k} \alpha_z(w) d_2(p_z \| \bfal'\bfp)p^\infty_W(w)dw\\
&\stackrel{(a)}{\geq}& \frac{2}{k\ln 2} \int_0^1 \sum_{z=0}^{k} \alpha_z(w) (p_z-\bfal'\bfp)^2 p^\infty_W(w)dw\\
&\stackrel{(b)}{\geq}& \frac{2}{k\ln 2} \frac{\left[\int_0^1 \sum_{z=0}^{k} f_1(z,w) \frac{1}{\sqrt{w(1-w)}}p^\infty_W(w)dw\right]^2}{\int_0^1 \sum_{z=0}^{k} f_2(z,w) \frac{1}{w(1-w)}p^\infty_W(w)dw}\\
&\stackrel{(c)}{=}& \frac{2}{k\ln 2} \frac{\left[\frac{1}{\pi} \int_0^1 (\frac{\partial \bfal}{\partial w})'\bfp dw\right]^2}{k}\\
&\stackrel{(d)}{=}& \frac{2}{k^2 \pi^2\ln 2},
\end{eqnarray*}
where 
$$f_1(z,w) = \alpha_z(w) (p_z-\bfal'\bfp)(z-kw)$$
and 
$$f_2(z,w) = \alpha_z(w) (z-kw)^2.$$
(a) follows from Pinsker's inequality \cite[Lemma
11.6.1]{Cover2006}. (b) follows from the Cauchy-Schwarz
inequality. The numerator of (c) follows from 
\begin{eqnarray*}
\sum_{z = 0}^k f_1(z,w) &=& \sum_{z=0}^{k} \alpha_z(w) (z-kw)p_z \\
&& \quad -\bfal'\bfp \underbrace{\bbE\left[Z-kw|W=w\right]}_{=0} \\
&=& w(1-w)(\frac{\partial \bfal}{\partial w})'\bfp,
\end{eqnarray*}
while the denominator follows from
$$\sum_{z = 0}^k f_2(z,w) = \bbE\left[(Z-kw)^2|W=w\right] = kw(1-w).$$
Finally, (d) follows from the marking assumption: $\bfal'(0)\bfp = 0$ and $\bfal'(1)\bfp = 1$.
\end{proof}

\subsection{Asymptotic Behavior for Large Coalition} \label{subsec:asym}

\begin{figure}
\centering
\includegraphics[width=3.2in]{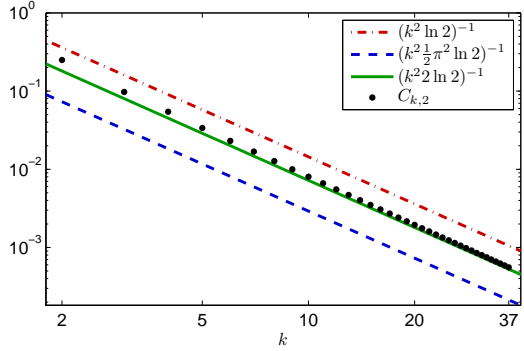}
\caption{Capacity $C_{k,2}$ and upper and lower bounds}
\label{fig:C_plot}
\end{figure}

\begin{figure}
\centering
\includegraphics[width=3.2in]{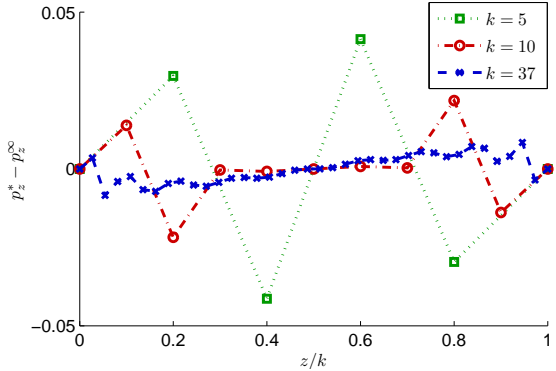}
\caption{Difference $p^*_z - p^\infty_z$ for $k = 5$, $10$, and $37$}
\label{fig:p_plot}
\end{figure}

\begin{figure}[t]
\centering
\includegraphics[width=3.2in]{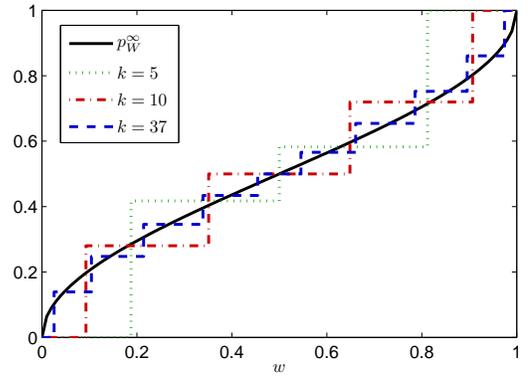}
\caption{Cumulative distribution function of $p_W^\infty$ and $p_W^*$ for $k = 5$, $10$, and $37$}
\label{fig:cdf_plot}
\end{figure}

We solve the capacity games for small $k$'s using \eqref{eqn:dcdw} and \eqref{eqn:dcdp} in Sec. \ref{subsec:analysis}. Fig. \ref{fig:C_plot} shows the capacity along with the upper and lower bounds. Amiri and Tardos \cite{Amiri2009} stated without proof that $C_{k,2} \gtrsim (k^2 2\ln2)^{-1}$. Our numerical results suggest that this bound is tight and that the convergence is fairly quick.

Evaluating the convex game of \eqref{eqn:C_k2_maxmin} or \eqref{eqn:C_k2_minmax} for large $k$ is still a difficult task. However, Theorem \ref{thm:ub} and \ref{thm:lb} shed lights on the asymptotic behavior of the game. If a less powerful coalition simply chooses the interleaving attack as their strategy (a.k.a. ``uniform channel'' in \cite{Anthapadmanabhan2008} and ``blind colluders'' in \cite{Furon2008}), Theorem \ref{thm:ub} shows that the gain in rate is no more than a factor of two. In fact, one can show that $\hat{C}(w,\bfp^\infty) \sim (k 2\ln2)^{-1}$ for all $w \in (0,1)$ (based on results from \cite[\S1.6]{Lorentz1986}). Fig. \ref{fig:p_plot} shows the difference between $\bfp^*$ and $\bfp^\infty$ for different values of $k$. This suggests that the interleaving attack is asymptotically optimal. This also answers Furon {\it et al.}'s question in \cite{Furon2008}: the fingerprinting code can only be slightly shorter even against a naive coalition who performs solely the interleaving attack.

An even bigger issue for the content distributor is that the computation of the optimal $p_W^*$ is infeasible for large $k$. Luckily, Theorem \ref{thm:lb} resolves this predicament. By using $p_W^\infty$ of \eqref{eqn:pwinfty}, the loss in rate is only by a factor of about 2.5. Fig. \ref{fig:cdf_plot} suggests, surprisingly, that $p_W^*$ converges to $p_W^\infty$ in distribution. The same distribution was used in Tardos' fingerprinting code in \cite{Tardos2003}, which uses a \emph{simple} decoder and is designed to be independent of the collusion channel \cite{Furon2008}. This unifies the asymptotic distribution of $p_W^*$ for the simple and \emph{joint} decoders (see \cite{Moulin2008a}: $p_W^\infty$ is asymptotically optimal.

We conclude with the following conjecture:

\begin{conjecture}
When $k \rightarrow \infty$, we have 
\begin{equation}
C_{k,2} \sim (k^2 2\ln 2)^{-1},
\end{equation}
\begin{equation}
\bfp^* \sim \bfp^\infty,
\end{equation}
and
\begin{equation}
p_W^* \rightarrow p_W^\infty \textrm{ in distribution.}
\end{equation}
\end{conjecture}

\section*{Acknowledgment}

The authors would like to thank N. Prasanth \mbox{Anthapadmanabhan} for illuminating discussions and helpful comments.

This research is supported by NSF under grants CCF 06-35137 and CCF 07-29061.

\bibliographystyle{IEEEtran}
\bibliography{IEEEabrv,ref}

\end{document}